\documentclass[aps,prb,reprint,showpacs,floatfix,letterpaper,longbibliography]{revtex4-2}
\usepackage{graphicx}
\usepackage[english]{babel}
\usepackage{bm}
\usepackage{amsmath,amsfonts}
\usepackage{amsbsy}
\usepackage{dsfont}
\usepackage[colorlinks=true,linkcolor=blue,urlcolor=blue,citecolor=blue,breaklinks=true]{hyperref}
\usepackage[utf8]{inputenc}
\DeclareUnicodeCharacter{03BC}{\mbox{$\mu$}}
\DeclareUnicodeCharacter{2009}{ }

\usepackage{graphics}
\usepackage{subcaption}
\usepackage{xcolor}
\usepackage{bbm}
\def\bm#1{\mbox{\boldmath{$#1$}}}

\newcommand{\be}{\begin{equation}}
\newcommand{\ee}{\end{equation}}
\newcommand{\backvv}[1]{\reflectbox{$\vv{\reflectbox{\!$#1$}}$}}

\usepackage[normalem]{ulem}
\usepackage{mathtools} 
\usepackage{hyperref}
\usepackage{stmaryrd}
\usepackage{float}
\usepackage{physics}
\usepackage{dsfont}
\usepackage{esvect}
\hypersetup{colorlinks,allcolors=blue}
\usepackage{amsmath}
\usepackage{tcolorbox}
\usepackage{graphicx}
\usepackage{amssymb} 
\usepackage{amsthm} 
\usepackage{tikz}
\usepackage{graphicx}
\usepackage{comment}
\usepackage[justification=raggedright,singlelinecheck=false]{caption}

\begin{document}

\title{I-V characteristics of SNS junctions with a multivalley 
normal region
}
\author{Liam Bonds}
\affiliation{Department of Physics, University of Washington, Seattle, Washington 98195, USA}
\author{Shiang-Bin Chiu}
\affiliation{Department of Physics, University of Washington, Seattle, Washington 98195, USA}
\author{Anton Andreev}
\affiliation{Department of Physics, University of Washington, Seattle, Washington 98195, USA}
\author{Boris Spivak}
\affiliation{Department of Physics, University of Washington, Seattle, Washington 98195, USA}
\date{\today}

\begin{abstract}
In multivalley conductors the inter-valley relaxation time $\tau_v$  
and the inelastic relaxation time $\tau_{in}$
may be significantly longer than the intra-valley momentum relaxation time $\tau$. 
We show that this separation of time scales has dramatic effects on the I-V characteristics of SNS junctions with a multivalley normal region. We generalize the Larkin-Ovchinnikov equations describing superconducting kinetics to the case of multivalley superconductors. We use this generalization to obtain 
a kinetic description of multivalley SNS junctions. 
We find that at constant voltage bias $V$, the current $I(V)$ is nonmonotonic; it exhibits two peaks of similar magnitude $I_\text{max,1} \sim I_\text{max,2}$ at  $V_1 \sim \hbar(e\tau_{in})^{-1}$ and $V_2\sim \hbar(e\tau_v)^{-1}$, which may greatly exceed the critical current $I_c(T)$. At constant current bias $I$ we find that in a wide interval, $I_c(T) < I \lesssim I_{\text{jump}}$, the nonlinear resistance of the junction is controlled by the long relaxation times and may be several orders of magnitude smaller than the normal state resistance. 
\end{abstract}

\maketitle

\section{Introduction}

In many conductors, the Fermi surfaces  consist of several disconnected components, which surround electron and/or hole pockets. In particular, multi-component Fermi surfaces are  realized in doped semiconductors with multivalley electron dispersion of the valence or conduction bands. The inter-valley relaxation time $\tau_{v}$ may be much longer than the intra-valley momentum relaxation time $\tau$. 
In lightly doped  semiconductors with a multivalley dispersion, this occurs because the inter-valley electron scattering involves the transfer of momentum of the order of the size of the Brillouin zone, whereas the Fermi momentum within a given valley may be  relatively small. This hierarchy of relaxation times results in several new effects in the transport properties of semiconductors (see for example Refs. \cite{blount1959ultrasonic,adler1964velocity,gantsevich1967theory,rashba1976anisotropic}).

Recently, much experimental research was focused on composite superconductor-semiconductor structures 
\cite{takayanagi1985superconducting,kleinsasser1989superconducting,akazaki1996josephson,giazotto2004josephson,shabani2016two,kjaergaard2016quantized,kjaergaard2017transparent,bottcher2018superconducting,nishino1986carrier,hatano1988influence,chiodi2017proximity,hendrickx2019ballistic,aggarwal2021enhancement,tosato2023hard,valentini2024parity}. 
In the case of superconductor-silicon (s-Si), \cite{nishino1986carrier,hatano1988influence,chiodi2017proximity}, or superconductor-germanium (s-Ge), \cite{hendrickx2019ballistic,aggarwal2021enhancement,tosato2023hard,valentini2024parity} structures, the proximitized normal region may host
multivalley Fermi surfaces.

\begin{figure}
    \centering    \includegraphics[width=1\linewidth]{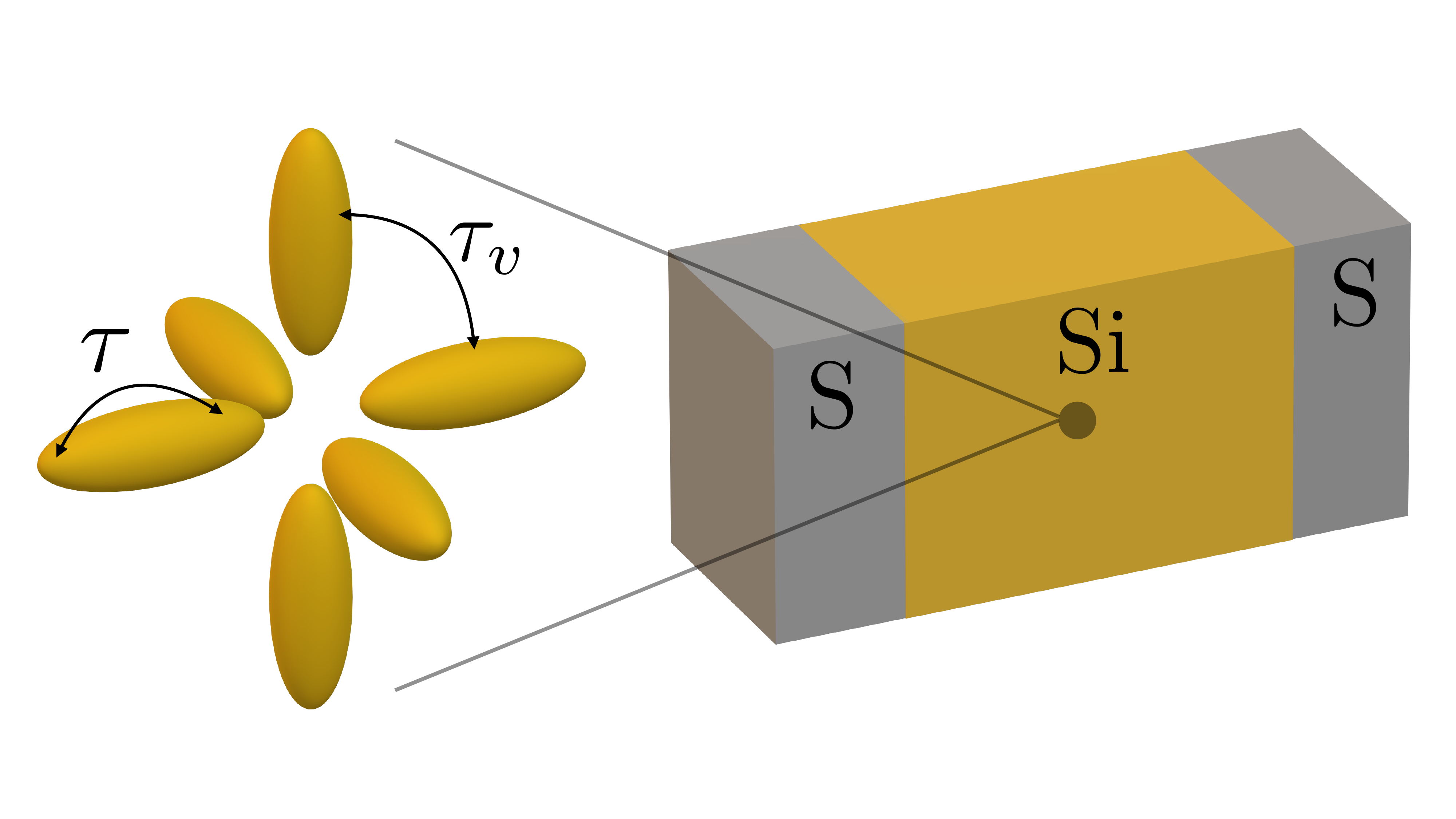}
    \caption{Sketch of a multivalley SNS junction with an n-doped silicon (Si) normal region. The orientation of  the six Fermi pockets relative to the device geometry is shown at left. The double-arrowed arcs indicate the scattering processes associated with the relaxation times $\tau$ and $\tau_v$. }
    \label{fig:deviceGeometry}
\end{figure}

In this article we generalize the Larkin-Ovchinnikov equations describing superconducting kinetics to the case of multivalley superconductors. We then use  these equations to develop a theory of the current-voltage (I-V) characteristics of superconductor-normal metal-superconductor junctions with a multivalley normal region illustrated in Fig.~\ref{fig:deviceGeometry}.

In the single valley case, the theory of I-V characteristics of  superconducting weak links and SNS junctions at relatively large voltages has been developed in many articles (see for example \cite{larkin1967tunnel,tinkham2004introduction,artemenko1979theory,averin1996adiabatic},  and references therein).
The main feature of the I-V characteristics at large voltages is that, similarly to the linear conductivity of normal metals, it is controlled by the shortest relaxation time in the system, $\tau$.

On the other hand, the small voltage regime attracted much less attention. It has been shown that in the single valley case the low voltage the I-V characteristics of SNS junctions  is controlled by the longest relaxation time in the system, the inelastic relaxation time $\tau_{in}$~\cite{liu2023current,liu2024giant_nonreciprocity}. SNS junctions with a multivalley normal region
are characterized by two long relaxation time scales,  $\tau_{in}$ and $\tau_{v}$. In this article we show that
the hierarchy of the relaxation times $\tau_{in}\gg \tau_{v}\gg \tau$ results in new features of the low voltage part of the I-V characteristics, as compared to the single valley case.

The presentation is organized as follows. In Sec.~\ref{sec:general} we generalize the Larkin-Ovchinnikov (LO) equations to diffusive multivalley  superconductors. In Sec.~\ref{sec:SNS} we apply the generalized LO equations to develop a theory of multivalley SNS junctions. 

\begin{widetext}
\section{Description of multivalley superconductors in the diffusive regime\label{sec:general}}

In  multivalley superconductors, in addition to the space-time position $\mathbf{X} = (t,\mathbf{x})$ and the spin index $\uparrow/\downarrow$, the
electron creation and annihilation  operators, $\psi^\dagger$ and $\psi$, must be labeled  by the valley index $\alpha$. We consider conventional $s$-wave superconductors, in which the Cooper pairing occurs between states related by time-reversal (TR) symmetry, and denote the TR partner of valley $\alpha$ by $-\alpha$. Therefore, the electron operator in the Gor'kov-Nambu space has the form
\begin{equation}
\Psi^\alpha (\mathbf{X}) =
   \begingroup
\renewcommand*{\arraystretch}{1.3}
\begin{pmatrix}
\psi^\alpha_{\uparrow} (\mathbf{x},t)\\
    \psi^{-\alpha,\dag}_{\downarrow} (\mathbf{x},t)
    \end{pmatrix}.
\endgroup
\end{equation}
The retarded, advanced, and Keldysh Green's functions are defined by, 
\begin{subequations}
\label{eq:G_Psi_def}
\begin{align}\label{eq:G_R_Psi_def}
 \hat{G}^{\alpha\beta}_R (\mathbf{X},\mathbf{X}') =&  - i \Theta (t-t')\left<\!\!\!\!\!\!\left<\begin{pmatrix}
     \{ \psi^\alpha_{\uparrow} (\mathbf{X}), \psi^{\beta,\dagger}_{\uparrow}(\mathbf{X}')\} &  \{ \psi^\alpha_{\uparrow} (\mathbf{X}), \psi^{-\beta}_\downarrow(\mathbf{X}')\}\\
      -\{ \psi^{-\alpha,\dag}_{\downarrow} (\mathbf{X}), \psi^{\beta,\dagger}_{\uparrow}(\mathbf{X}')\} &  -\{ \psi^{-\alpha,\dag}_{\downarrow} (\mathbf{X}), \psi^{-\beta}_\downarrow(\mathbf{X}')\}\end{pmatrix}\right>\!\!\!\!\!\!\right>,\\
\label{eq:G_A_Psi_def}
\hat{G}^{\alpha\beta}_A (\mathbf{X},\mathbf{X}') =& \ i \Theta (t'-t) \left<\!\!\!\!\!\!\left<\begin{pmatrix}
     \{ \psi^\alpha_{\uparrow} (\mathbf{X}), \psi^{\beta,\dagger}_{\uparrow}(\mathbf{X}')\} &  \{ \psi^\alpha_{\uparrow} (\mathbf{X}), \psi^{-\beta}_\downarrow(\mathbf{X}')\}\\
      -\{ \psi^{-\alpha,\dag}_{\downarrow} (\mathbf{X}), \psi^{\beta,\dagger}_{\uparrow}(\mathbf{X}')\} &  -\{ \psi^{-\alpha,\dag}_{\downarrow} (\mathbf{X}), \psi^{-\beta}_\downarrow(\mathbf{X}')\}\end{pmatrix}\right>\!\!\!\!\!\!\right>,\\
    \label{eq:G_K_Psi_def}
\hat{G}^{\alpha\beta}_K (\mathbf{X},\mathbf{X}') =&  -i\left<\!\!\!\!\!\!\left< \begin{pmatrix}
     [ \psi^\alpha_{\uparrow} (\mathbf{X}), \psi^{\beta,\dagger}_{\uparrow}(\mathbf{X}')] &  [\psi^\alpha_{\uparrow} (\mathbf{X}), \psi^{-\beta}_\downarrow(\mathbf{X}')]\\
      -[ \psi^{-\alpha,\dag}_{\downarrow} (\mathbf{X}), \psi^{\beta,\dagger}_{\uparrow}(\mathbf{X}')] &  -[ \psi^{-\alpha,\dag}_{\downarrow} (\mathbf{X}), \psi^{-\beta}_\downarrow(\mathbf{X}')]\end{pmatrix}\right>\!\!\!\!\!\!\right>.
\end{align}
\end{subequations}
Here  $[A , B ]$ and $\{ A, B\}$ denote, respectively, the commutator and anticommutator of $A$ and $B$. The expectation value  $\left<\!\left<\cdot\right>\!\right>
$ is with respect to the dynamic state of the system. We denote the matrices in the Gor'kov-Nambu space by a hat, e.g. $\hat{G}^{\alpha\beta}_R$. The three Green's functions in Eq.~\eqref{eq:G_Psi_def} may be combined into a single matrix in the Keldysh space, 
\begin{equation}\label{eq:G-def}
    \check G^{\alpha \beta} = \begin{pmatrix}
        \hat G_R^{\alpha\beta} & \hat G^{\alpha\beta}_K \\
        0 & \hat G^{\alpha\beta}_A
    \end{pmatrix}.
\end{equation}
The matrices in the Keldysh-Gor'kov-Nambu space are denoted by a check,  $\check{G}^{\alpha\beta}$.

To describe the intra-valley and inter-valley elastic scattering of electrons, we introduce two types of disorder potentials into the Hamiltonian:
1) the valley-diagonal disorder $U^\alpha$, which controls the intra-valley momentum relaxation, and 2) the valley-nondiagonal disorder $v^{\alpha \beta}$, which mediates elastic inter-valley scattering of electrons. The intra-valley momentum relaxation rate, and the inter-valley relaxation
rate are proportional to the variance of the disorder potentials, see  Eq.~\eqref{eq:variances}.
For a given realization of disorder, the electron Green's function $(\check{G}^{-1})^{\alpha \beta}$ is defined by,

\begin{equation}
\label{eq:G_inverse_def}
\left(\check{G}^{-1} \right)^{\alpha \beta} = i\hbar\check{\tau}_3\partial_t -  \left[\check{H}^{\alpha}_{\text{BCS}} + U^\alpha (\mathbf{x}) \right]\delta^{\alpha\beta} -v^{\alpha\beta}(\mathbf{x})  -\check{\Sigma}^{\alpha\beta}_{\mathrm{in}}.
\end{equation} 
Here $\check{\tau}_i\equiv \hat{\tau}_i\otimes \mathds{1}_K$ acts as the Pauli matrices $\hat{\tau}_i$ in the Gor'kov-Nambu space and as the identity ${\mathds{1}}_K$ in the Keldysh space. The inelastic self-energy, $\check{\Sigma}^{\alpha\beta}_{\text{in}}$, describes the inelastic scattering processes, and the BCS mean-field Hamiltonian, $\check{H}^{\alpha}_{\text{BCS}}\equiv \hat{H}^{\alpha}_{\text{BCS}}\otimes \mathds{1}_{K}$, is diagonal in the valley indices, and takes the form
\begin{equation}
\label{eq:H_BdG}
\hat{H}^{\alpha}_{\text{BCS}}=    -\left[ 
\frac{\hbar^2}{2}\left(m^{\alpha}\right)^{-1}_{ij}\left(\nabla_i - i\frac{e}{\hbar c} A_i \hat{\tau}_3 \right) \left(\nabla_j - i\frac{e}{\hbar c} A_j \hat{\tau}_3 \right) +\mu^\alpha - e\phi  \right]  -\hat{\Delta}^\alpha.
\end{equation}
In this equation $i,j$ are the Cartesian indices, $\left(m^{\alpha}\right)^{-1}_{ij}$ is the inverse effective mass tensor in valley $\alpha$, and $(\phi,A_i)$ are the scalar and vector potentials. The combination $\mu^\alpha - e\phi $ is the electrochemical potential in valley $\alpha$, measured from the bottom of the band in that valley. The energy of the bottom of the conduction band may depend on the valley index $\alpha$ because of strain or because the valleys are not related by a crystal symmetry.  As a result, the density of states at the Fermi level $\nu_N^\alpha$ is in general valley-dependent.  
The order parameter, $\hat \Delta^\alpha =  i\Delta^\alpha\hat{\tau}_2 $, is symmetric between the pair of valleys $(\alpha,-\alpha)$, but may differ between different TR-related pairs. For the remainder of the article we will work in units $\hbar=c=1$.

The Larkin-Ovchinnikov equations~\cite{larkin1977non} describe the quasiparticle kinetics in the diffusive regime, in which the characteristic time scales exceed the elastic momentum relaxation time $\tau$, and the characteristic length scales exceed the elastic mean free path $\ell \sim v_F \tau$, where $v_F$ is the Fermi velocity. These  equations are 
written in terms of the disorder-averaged Green's functions at coinciding spatial points. In multivalley
superconductors, the averaging is performed over two types of disorder and is characterized by two relaxation times. 
We assume that the inter-valley relaxation rate is small, $1/\tau_v \ll 1/\tau$. The derivation proceeds in the standard way, as outlined 
in Appendix~\ref{Appendix}.

The disorder-averaged Green's functions are diagonal in the valley index $\alpha$. 
Denoting the diagonal components of the Green's functions Eqs.~(\ref{eq:G_R_Psi_def}-\ref{eq:G_K_Psi_def}) by $\check{{G}}^\alpha$ we can write the disorder-averaged Green's functions at coinciding points  in the form,
\begin{equation}
\check{g}^\alpha\left(\epsilon,t,\mathbf{r}\right) = \frac{ i}{\pi\nu_N^\alpha}\int dt' e^{i\epsilon t'} \left\langle\check{G}^\alpha(t-t'/2,\mathbf{r},t+t'/2,\mathbf{r})\right\rangle,\label{eq:quasiclassicalgf_average}
\end{equation} 
where $\langle \ldots \rangle$ denotes disorder averaging, and the normal state density of states at the Fermi level, $\nu_N^\alpha$, includes spin degeneracy.
The dimensionless quasiclassical Green's functions $\check{g}^\alpha$ satisfy the standard nonlinear constraint~\cite{larkin1977non},
\begin{equation}\label{eq:nonlinearconstraint}
    (\check{g}^\alpha)^2=1.
\end{equation}
Here and below, we suppress the obvious arguments of the semiclassical Green's functions for the sake of brevity.

The retarded and advanced Green's functions obey the generalized Usadel equation,
\begin{equation}\label{eq:multiUsadel}
D^\alpha_{ij}\left[\partial_i\left(\hat{g}^\alpha_R\partial_j\hat{g}^\alpha_R \right)\right] + i\left[\epsilon\hat{\tau}_3+\hat{\Delta}^\alpha,\hat{g}^\alpha_R\right] =  \sum_{\beta\neq\alpha} \frac{1}{2\tau_{\alpha\beta}}\left[\hat{g}^\beta_R,\hat{g}^\alpha_R\right].
\end{equation}
Here $D_{ij}^\alpha$ is the diffusion tensor in valley $\alpha$ and summation over the repeated Cartesian indices $(i,j)$ is implied. The covariant derivative is, $\partial_j = \nabla_j - ie A_j\left[\hat{\tau}_3,\cdot\right]$. The left-hand side of Eq.~\eqref{eq:multiUsadel} corresponds to the standard Usadel equation in an individual valley $\alpha$. The right-hand side describes the coupling of the retarded Green's functions in different valleys caused by elastic inter-valley scattering. The inter-valley relaxation rate $\tau_{\alpha\beta}^{-1}$  is given by Eq.~\eqref{eq:variances}.

The kinetic equations for the quasiparticle distribution functions are obtained in the standard way by writing the time-evolution equations for the Keldysh component of the disorder-averaged Green's function at coinciding points. 
The latter contains information about both the quasiparticle states and their occupancy. In order to separate the evolution of the quasiparticle occupation from that of the spectrum, different parameterizations have been suggested, see Refs.~\cite{nielsen1985thermal,shelankov1985derivation}. Below, we use the Larkin-Ovchinnikov parametrization,
\begin{equation}
\label{eq:G_K_f}
    \hat{g}_K^\alpha = \hat{g}_R^\alpha\hat{h}^\alpha - \hat{h}^\alpha\hat{g}_A^\alpha, \quad \hat{h}^\alpha=f^\alpha_0\hat{\tau}_0+f^\alpha_1\hat{\tau}_3.
\end{equation}
We note that $f^\alpha_1(\epsilon)$ describes the particle-hole branch imbalance in a nonequilibrium state. In particular, it describes the current in the normal state (see corresponding discussion in~\footnote{The physical meaning of the two distribution functions may be understood by considering  relatively clean superconductors, with $\Delta \tau \gg 1$. In this case, one can use the Boltzmann equation for the quasiparticle distribution function $n^\alpha ({\bf p})$~\protect\cite{aronov1981boltzmann}. In the diffusive regime, the
distribution function becomes isotropic $n^\alpha ({\bf p}) \to  n^\alpha(\xi_{\bf p})$. The quasiparticle energy is $\epsilon=\sqrt{|\Delta|^2 + \xi_{\bf p}^2}$, and the functions
$f^\alpha_1(\epsilon)$ and $f^\alpha_0(\epsilon)$  
are given by
$f_1^{\alpha}(\epsilon) =  \frac{\xi_{\bf p}}{\epsilon_{\bf p}}\left[ n^\alpha(\xi_{\bf p}) - n^\alpha(-\xi_{\bf p}) \right]$, and 
 $f_0^{\alpha}(\epsilon)= 1-n^\alpha(\xi_{\bf p}) - n^\alpha(-\xi_{\bf p})$. In equilibrium we have $f_1^\alpha(\epsilon)=0$ and $f_0^\alpha (\epsilon) = \tanh (\epsilon/2T)$. 
 }).

With the parametrization in Eq.~\eqref{eq:G_K_f}, the time-evolution equation for the Keldysh Green's functions reduces to a system of kinetic equations for the $f_0^\alpha$ and $f_1^\alpha$ distribution functions,
\begin{align}
\label{eq:kineq0}
      \nonumber \nu^\alpha\partial_tf^\alpha_0 -\frac{\nu^\alpha_N}{4}D_{ij}^\alpha\bigg[\partial_i\left(\mathcal{K}_0^\alpha \partial_jf^\alpha_0- \mathcal{Y}^\alpha\partial_jf^\alpha_1\right) +j^\alpha_j\partial_if^\alpha_1-e \partial_t A_i[\partial_\epsilon\left(\mathcal{Y}^\alpha\partial_j 
      f^\alpha_0\right)+\partial_\epsilon\left(\mathcal{K}_1^\alpha\partial_jf^\alpha_1\right)]\bigg]& \\    -\partial_\epsilon f^\alpha_0\int^\epsilon_0 (\partial_t \nu^\alpha) d \epsilon'- \frac{i\nu^\alpha_N}{2}f^\alpha_1\Tr\left[\hat{\Delta}^\alpha\hat{\tau
     }_3\left(\hat{g}^\alpha_R-\hat{g}^\alpha_A\right)\right] +  \frac{\nu^\alpha_N}{4}\Tr \left[\hat \Delta^\alpha \hat \tau_3 \partial_\epsilon \left(\hat{g}^\alpha_R + \hat{g}^\alpha_A\right)\right]\partial_t f^\alpha_1 &= \text{St}[f^\alpha_0]
\end{align}
and
\begin{align}\label{eq:kineq1}
     \nonumber \nu^\alpha\partial_tf^\alpha_1 -\frac{\nu^\alpha_N}{4}D_{ij}^\alpha\bigg[\partial_i\left(\mathcal{K}_1^\alpha\partial_jf^\alpha_1+ \mathcal{Y}^\alpha\partial_jf^\alpha_0\right)+j^\alpha_j\partial_if^\alpha_0-e \partial_t A_i\left[j^\alpha_j\partial_\epsilon f^\alpha_1-\partial_\epsilon\left(\mathcal{Y}^\alpha\partial_jf^\alpha_1\right)+\partial_\epsilon\left(\mathcal{K}_0^\alpha\partial_jf^\alpha_0\right)\right]\bigg]&\\  - \frac{i\nu^\alpha_N}{2}f^\alpha_1\Tr\left[\hat{\Delta}^\alpha\left(\hat{g}^\alpha_R+\hat{g}^\alpha_A\right)\right]+\nonumber \frac{\nu^\alpha_N}{4}\Tr\left[\hat{\tau}_3\partial_t\hat{\Delta}^\alpha\left(\hat{g}^\alpha_R+\hat{g}^\alpha_A\right)\right]\partial_\epsilon f^\alpha_0&\\+ \frac{\nu^\alpha_N}{4}\Tr[\hat{\Delta}^\alpha\partial_\epsilon\left(\hat{g}^\alpha_R-\hat{g}^\alpha_A\right)]\partial_tf^\alpha_1+ \nu^\alpha \partial_t(e\phi-\mu^\alpha)\partial_\epsilon f_0^\alpha&=\text{St}[f^\alpha_1].
\end{align}
Here we defined the coherence factor $\mathcal{Y}^\alpha$, and the dimensionless diffusion coefficients $\mathcal{K}^\alpha_0$ and $\mathcal{K}_1^\alpha$, as 
\begin{equation}\label{eq:diff+coherencefactors}
    \mathcal{Y}^\alpha =\Tr\left[\hat{\tau}
    _3\hat{g}_A^\alpha\hat{g}_R^\alpha\right],\quad
    \mathcal{K}^\alpha_0 = \Tr\left[1-\hat{g}_R^\alpha \hat{g}_A^\alpha\right], \quad
    \mathcal{K}_1^\alpha = \Tr\left[1-\hat{\tau}_3\hat{g}_R^\alpha \hat{\tau}_3\hat{g}_A^\alpha\right].
\end{equation}
The quasiparticle density of states $\nu^\alpha$ and the spectral current density $j^\alpha_j$ in valley $\alpha$ are expressed in terms of the retarded Green's functions as,
\begin{equation}\label{eq:dosDefinition}
    \nu^\alpha = \frac{\nu_N^\alpha}{2}\Re\left\{\Tr\left[\hat{\tau}_3\hat{g}_R^\alpha\right]\right\}, \quad j^\alpha_j = \Tr\left[\hat{\tau}_3\left(\hat{g}_R^\alpha\partial_j\hat{g}_R^\alpha - \hat{g}_A^\alpha\partial_j\hat{g}_A^\alpha\right)\right].
\end{equation}
Finally the collision integrals have the form, 
\begin{subequations}
\label{eq:CollisionIntegrals}
\begin{align}\label{eq:CollisionIntegral_0}
   \text{St}[f^\alpha_0] = &\, -\sum_{\beta}\frac{\nu^\alpha_N}{8\tau_{\alpha\beta}} \Tr\bigg(\hat{\tau}_3\left[\left(\hat{g}_R^\alpha-\hat{g}_A^\alpha\right)\hat{g}_R^\beta-\hat{g}_A^\beta\left(\hat{g}_R^\alpha-\hat{g}_A^\alpha\right)\right]\left(f_1^\alpha - f_1^\beta\right)+\left(\hat{g}_R^\alpha-\hat{g}_A^\alpha\right)\left(\hat{g}_R^\beta-\hat{g}_A^\beta\right)\left(f_0^\alpha-f_0^\beta\right)\bigg) \nonumber \\
   &    -\frac{\nu^\alpha(f_0^\alpha - f_F)}{\tau_{in}},\\
\label{eq:CollisionIntegral_1}
    \text{St}[f_1^\alpha] = &\,  -\sum_{\beta}\frac{\nu^\alpha_N}{8\tau_{\alpha\beta}} \Tr\bigg(\hat{\tau}_3\left[\hat{g}_R^\alpha\left(\hat{g}_R^\beta-\hat{g}_A^\beta\right)-\left(\hat{g}_R^\beta-\hat{g}_A^\beta\right)\hat{g}_A^\alpha\right]\left(f_0^\alpha - f_0^\beta\right)+ \left(\hat{g}_R^\alpha\hat{\tau}_3-\hat{\tau}_3\hat{g}_A^\alpha\right)\left(\hat{\tau}_3\hat{g}_R^\beta-\hat{g}_A^\beta\hat{\tau}_3\right)\left(f_1^\alpha-f_1^\beta\right)\bigg) \nonumber \\
    & -\frac{\nu^\alpha f_1^\alpha }{\tau_{in}}.
\end{align}
\end{subequations}
We wrote the inelastic collision integral in the relaxation time approximation, where 
\begin{align}
\label{eq:f_Fermi}
    f_F(\epsilon) = & \, \tanh\left(\frac{\epsilon}{2T}\right)
\end{align}
is the equilibrium distribution function.

We note that the kinetic equation Eq.~\eqref{eq:kineq0} differs from the standard form, see Ref.~\cite{larkin1977non}. To arrive at this form, we 
used the identity
\begin{equation}
      \partial_t\nu^\alpha + \frac{\nu^\alpha_N}{4}\partial_\epsilon \left[ D^\alpha_{ij}(e\partial_t A_i) j^\alpha_j +\Tr\left[\partial_t \hat \Delta^\alpha (\hat g_R^\alpha-\hat g_A^\alpha)\right]\right]= 0,
    \label{eq:conservation dos}
\end{equation}
which is derived in the Appendix, see the discussion below Eq.~\eqref{LarkinOvchinnikov}. This identity expresses conservation of the number of quasiparticle states and has the form of a continuity equation. The second term in Eq.~\eqref{eq:conservation dos} is the divergence (in energy space) of spectral current which is caused by the time evolution of $A_i$ and $\Delta^\alpha$. Its single valley analogue appears in the LO kinetic equations (Eq.~(A.32) in Ref.~\cite{larkin1986nonequilibrium}). In Eq.~\eqref{eq:kineq0} the divergence of the spectral current was replaced with $\partial_t \nu^\alpha$ (the first term in Eq.~\eqref{eq:conservation dos}). The form of the kinetic equation given by Eq.~\eqref{eq:kineq0} is more convenient for analyzing the consequences of time-dependence of the quasiparticle density of states (spectral flow) on the kinetics of superconducting systems, which is carried out in Sec.~\ref{sec:SNS}.

\subsection{\label{sec:self-consistency}Expressions for the observables  
in terms of the distribution functions}

The kinetic equations \eqref{eq:kineq0} and \eqref{eq:kineq1} describe the evolution of the quasiparticle distribution functions in terms of the  parameters of 
the time-dependent BCS Hamiltonian Eq.~\eqref{eq:H_BdG}: $\Delta^\alpha$, $\mu^\alpha$, $\phi$, and $\mathbf{A}$. These parameters, in turn,   depend on the quasiparticle distribution functions. Therefore, the kinetic equations must be supplemented by a number of conditions, which relate the values of the self-consistent potentials $\Delta^\alpha$, $\mu^\alpha$, $\phi$, and $\mathbf{A}$ to the quasiparticle distribution functions. 
 
The scalar and vector potentials are related by the Maxwell's equations to the charge and current densities. The current density is given by the sum of current densities in each valley, which are expressed in terms of the quasiparticle distribution functions as~\footnote{For  $f^\alpha_1 =0$ and $f^\alpha_0(\epsilon)=f_F(\epsilon)$ Eq.~\eqref{eq:currentdensity} reproduces the expression for equilibrium supercurrent. We note that in a normal metal $\mathcal{K}_1=4$, so the expression for the current is $ \bm{J}^\alpha = -e\nu^\alpha_N D \intop_0^\infty d\epsilon \bm{\nabla} f^\alpha_1 = e\nu^\alpha_N D\intop_{-\infty}^\infty d\epsilon \bm{\nabla} \delta n^\alpha$. This is consistent with the correspondence of $f^\alpha_1$ to the particle-hole branch imbalance, see the footnote above Eq.~\eqref{eq:kineq0}.
},
\begin{equation}\label{eq:currentdensity}
    J^\alpha_i = -\frac{e \nu^\alpha_N D^\alpha_{ij}}{4} \int_0^\infty d\epsilon [j^\alpha_j f_0^\alpha +\mathcal{K}_1^\alpha\partial_j f^\alpha_1 + \mathcal{Y}^\alpha\partial_jf_0^\alpha].
\end{equation}
The deviation $\delta n^\alpha$ of the electron density in valley $\alpha$ from equilibrium 
is expressed in terms of the distribution function $f_1^\alpha$ as,
\begin{eqnarray}\label{eq:electron-density}
    \delta n^\alpha =  (e\phi+ \delta\mu^\alpha)\nu_N^\alpha+  \int_0^\infty d\epsilon (\nu^\alpha f_1^\alpha),
\end{eqnarray}
Here  $e \phi + \delta \mu^\alpha $ is the deviation of the electrochemical potential in valley $\alpha$ from equilibrium. 
The electric potential is related to the total charge density by the Poisson equation $\nabla^2  \phi = - 4\pi e \sum_\alpha \delta n^\alpha$.
In the regime of charge neutrality, the Poisson equation reduces to the condition~\footnote{In multivalley conductors, the charge neutrality condition Eq.~\eqref{eq:charge_neutrality} does not preclude deviations of the electron density $n^\alpha$ in a given valley from equilibrium.  It follows from Eqs.~\eqref{eq:electron-density} and \eqref{eq:charge_neutrality} that in the normal state, the electric field is related to $f^\alpha_1(\epsilon)$ by $ e \bm{E} = \frac{1}{\protect\sum_\alpha \nu_N^\alpha} \protect\sum_{\alpha} \intop_0^\infty \nu_N^\alpha \mathbf{\nabla} f_1^\alpha (\epsilon) d \epsilon$.}
\begin{align}
    \label{eq:charge_neutrality}
     \sum_\alpha \delta n^\alpha & \,  =0. 
\end{align}
Finally, the valley-dependent pair potential obeys the BCS self-consistency equation,
\begin{eqnarray}\label{eq:self-consistent}
     \Delta^\alpha
     = \frac{i}{4}\int_0^\infty d\epsilon\sum_{\beta} \lambda_{\alpha\beta} [(\hat g^\beta_R - \hat g^\beta_A)f_0^\beta +  (\hat g^\beta_R \hat\tau_3 - \hat\tau_3\hat g^\beta_A)f_1^\beta]_{12}.
\end{eqnarray}
Here $\lambda_{\alpha\beta}$ are the dimensionless BCS coupling constants for multi-band superconductors~\cite{Suhl_multiband}. They relate the pair potential $\Delta^\alpha$ in the pair of valleys $\alpha,-\alpha$  to the amplitude of the Cooper pair condensate in the pair of valleys $\beta, -\beta$. 
For the $s$-wave pairing considered here, $\lambda_{\alpha\beta}$ is invariant under $\alpha \to - \alpha$ and/or $\beta \to - \beta$, and 
$\Delta^\alpha =\Delta^{-\alpha}$. 

To summarize, the full system of equations describing the low-frequency kinetics in disordered multivalley superconductors consists of the Usadel equation, Eq.~\eqref{eq:multiUsadel}, the Larkin-Ovchinnikov equations for the quasiparticle distribution functions Eqs.~(\ref{eq:kineq0}-\ref{eq:kineq1}), the expressions for the charge and current densities Eqs.~(\ref{eq:currentdensity}-\ref{eq:electron-density}), the self-consistency equation for the pair potential Eq.~\eqref{eq:self-consistent}, and the Maxwell equations for the scalar and vector potential.

\end{widetext}
\section{Multivalley SNS junctions in the low bias regime}\label{sec:SNS}

In this section we apply the formalism of Sec.~\ref{sec:general} to study I-V characteristics of SNS junctions, in which the normal region has multivalley electron spectrum.  A schematic picture of an SNS junction 
sandwiched in between two $s$-wave superconductors,
is shown in Fig.~\ref{fig:deviceGeometry}.
In general, the proximity effect in the normal region of the junction is valley-dependent due to, e.g. the different orientation of the valleys relative to the N-S boundary of the device. We will assume that the diffusion coefficients in different valleys are of the same order of magnitude, $D^\alpha \sim D$, and 
that the normal region is in the diffusive regime, 
$L>\sqrt{D\tau}$, where $L$ is the width of the junction.

It was shown in Refs.~\cite{liu2023current,liu2024giant_nonreciprocity} that at low voltage, the I-V characteristics of SNS junctions with a single valley electron spectrum in the normal region are  controlled by the longest relaxation time in the system -- the inelastic relaxation time $\tau_{in}$. In multivalley junctions, the quasiparticle density of states becomes valley-dependent, and consequently the quasiparticle kinetics are characterized by two 
long relaxation times,  $\tau_{in}$ and $\tau_{v}$. Below, we consider the consequences of the existence of the two relaxation times on the I-V characteristics of multivalley junctions.

Qualitatively, the existence of the low bias regime where the nonlinear device conductance of the junction  significantly exceeds the normal state conductance,  can be understood as follows. The phase difference of the order parameter across the junction, $\chi(t)$, is related to the voltage $V(t)$  by the Josephson relation,
\begin{equation}\label{eq:Josephson}
\dot{\chi}(t)  = 2eV(t).
\end{equation} 
At low bias, $ e V \ll D/L^2$,  
the quasiparticle distribution function $f^\alpha_0$ becomes independent of position, and depends only on the energy $\epsilon$ and the valley index $\alpha$. In this regime, the quasiparticle distribution function is controlled entirely by the time evolution of the valley-dependent density of quasiparticle states integrated over the normal region,  $ \tilde{\nu}^\alpha(\chi(t)) = \int d\mathbf{r} \nu^\alpha({\bf r},\chi(t))$. The latter may be evaluated in the adiabatic approximation as a function of the instantaneous phase difference $\chi(t)$.
In a general situation,  the $\chi$-dependence of the densities of states  $\nu^{\alpha}( \chi(t))$ is different in different valleys.
Therefore, the voltage across the junction creates a nonequilibrium population of quasiparticles not only in energy but also in the valleys. 
We will focus on the regime $ L\ll \sqrt{D\tau_{v}}$, where the existence of a multivalley  spectrum in the normal region manifests itself in the most dramatic way. In this case the densities of states of individual valleys can be calculated independently.

Due to Andreev reflection from the normal metal-superconductor boundaries of the  SNS junction, the low-energy ($\epsilon< \Delta$) quasiparticles are trapped inside the normal region. Therefore, the nonequilibrium distribution of quasiparticles, which is generated by the time-dependence of the density of states, can relax only via the inter-valley and inelastic scattering. As a result, the conductance in the low bias regime turns out to be much greater than the normal state conductance~\cite{liu2023current}.

\begin{figure*}
    \centering
\includegraphics[width=.7\linewidth]{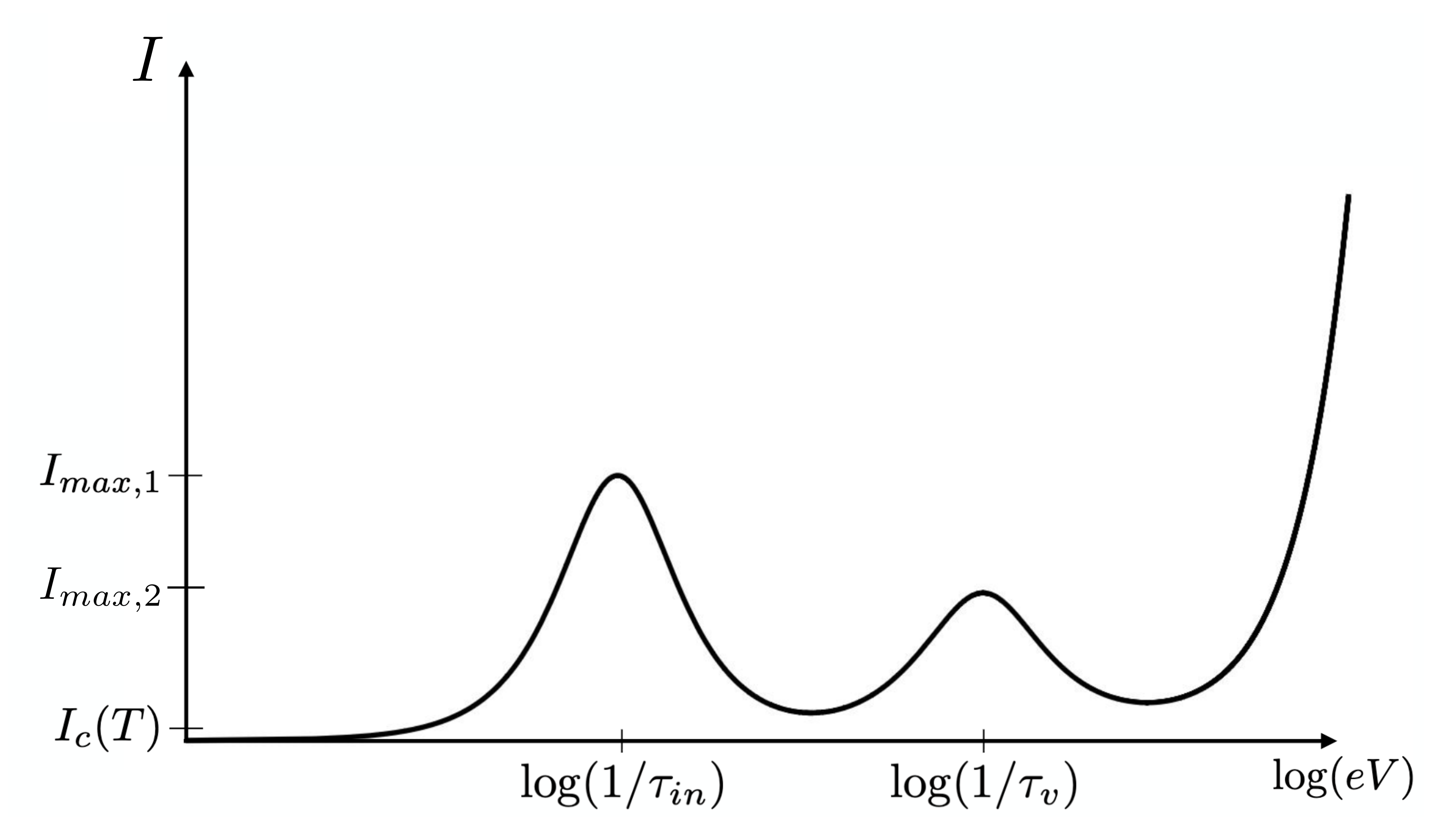}
    \caption{Plot of average current $I$ versus bias voltage $V$ in an SNS junction with a multivalley normal metal. At bias voltages of the order $eV_1 \sim \tau_{in}^{-1}$ and $ eV_2 \sim  \tau_v^{-1}$ there exist two maxima of the average current; $I_\text{max,1}, \  I_\text{max,2}$. These maxima are of the same order and may greatly exceed the temperature-dependent critical current $I_c(T)$.}\label{fig:VoltageBiasIV}
\end{figure*}

\subsection{Zero-dimensional description of multivalley SNS junctions at low bias\label{sec:zero_dimension}}

To develop a quantitative description of multivalley SNS junctions at low bias, we use the formalism presented in Sec.~\ref{sec:general}.  
For $eV \ll D/L^2$ the time-dependent quasiparticle distribution function inside the normal region becomes spatially-uniform, and depends only on the energy and the valley index $\alpha$. Thus, the problem becomes zero-dimensional. 
We assume that the temperature exceeds the characteristic value of the mini-gap in the normal region, $T\gtrsim D/L^2$. In this case, the dominant mechanism of dissipation is associated with the time-dependence of the density of quasiparticle states in the normal region. The evolution of the quasiparticle states creates nonequilibrium populations of quasiparticles in energies and valleys, which relax via inelastic and inter-valley scattering. 
The resulting dissipation rate turns out to be controlled by the long relaxation times, $\tau_v$ and $\tau_{in}$. For this reason the nonequilibrium effects described by the distribution function $f^\alpha_{1}$, whose relaxation is controlled by the intra-valley relaxation time $\tau$, may be neglected.

The zero-dimensional kinetic equation for the SNS junction in this regime 
is obtained by setting $f^\alpha_1=0$, integrating Eq.~\eqref{eq:kineq0} over space and dividing it by $\tilde \nu^\alpha(\epsilon, \chi)$. This yields (cf.~\cite{zhou1998density,liu2023current}),
\begin{equation}\label{EQ:N_DOT}
 \partial_{t} f^\alpha_0 (\epsilon, t)+ 2 e V(t) \mathds{V}^\alpha (\epsilon, \chi)\, \partial_\epsilon  f_0^\alpha(\epsilon, t) = \text{St}^\alpha_{\mathrm{v}}  + \text{St}^\alpha_\text{in}.
\end{equation}
Here $\mathds{V}^\alpha (\epsilon, \chi)$ is the sensitivity of the quasiparticle energy levels to changes of $\chi$.
It is given by 
\begin{equation}\label{eq:level_velocity}
\mathds{V}^\alpha (\epsilon,\chi) = \frac{-1}{\tilde \nu^\alpha (\epsilon, \chi)}  \int_{0}^{\epsilon} d \epsilon'  \frac{\partial \tilde \nu^\alpha (\epsilon', \chi)}{\partial \chi},
\end{equation}
where $\tilde \nu^\alpha(\epsilon, \chi)$ is the density of states integrated over space in valley $\alpha$,
\begin{equation}
\label{eq:global_DOS}
\tilde \nu^\alpha(\epsilon, \chi) =
\int_{V} d {\bf r} \nu^\alpha(\epsilon, {\bf r}, \chi).
\end{equation}
The inelastic and inter-valley collision integrals at $e V \ll E_T$,  may be described in the relaxation-time approximation, 
\begin{equation}\label{eq:RelaxTimeApprox}
\text{St}^{\alpha}_{v}= -\sum_{\beta} \frac{\delta f_0^{\alpha} -\delta f_0^{\beta}}{\tau_v}; \,\,\,\,\,\,\,\
  \text{St}^{\alpha}_{in}= - \frac{\delta f_0^{\alpha} }{\tau_{in}}.
\end{equation}
Here $ \delta f_0^\alpha(\epsilon,t) =f_0^\alpha(\epsilon,t)- f_F(\epsilon)$ is the nonequilibrium part of the distribution function, and we assumed that all inter-valley relaxation times are of the same order, $\tau_v$.

The current across the junction is given by 
\begin{align}
\label{EQ:CURRENTEUL}
     I= I_{c}(\chi,T)  + e  \sum_{\alpha} \int_{0}^{\infty} d \epsilon \ \tilde \nu^\alpha(\epsilon, \chi) \delta f^\alpha_0(\epsilon, t)\mathds{V}^\alpha(\epsilon, \chi),
\end{align}
where we absorbed the contribution of $f_F(\epsilon)$ into the equilibrium Josephson current $I_c(\chi,T)$ (first term in Eq.~\eqref{EQ:CURRENTEUL}). 

To derive Eq.~\eqref{EQ:CURRENTEUL} we applied the identity Eq.~\eqref{eq:conservation dos} to express the spectral current density $j_x^\alpha$, in Eq.~\eqref{eq:currentdensity}, in terms of the global density of states,
\begin{equation}\label{eq:jtoDOS}
      \frac{\nu^\alpha_N D^\alpha}{4L}\int_V  d\mathbf{r} j_x^\alpha(\epsilon,\mathbf{r},\chi) = - \int^{\epsilon}_0 d\epsilon' \frac{\partial \tilde\nu^\alpha(\epsilon',\chi)} {\partial \chi}. 
\end{equation}

Equations (\ref{EQ:N_DOT}-\ref{EQ:CURRENTEUL}) may be understood as a consequence of the conservation of the  number of quasiparticle levels in the normal region of the junction~\cite{giantMWAsmith2020,liu2024giant_nonreciprocity,smith2020giantSD,liu2023current}.
Integrating Eq.~\eqref{eq:conservation dos} 
over space, one obtains the continuity equation for the flux of quasiparticle states in energy space (spectral flow);
$
\partial_t \tilde{\nu}^{\alpha} + \partial_\epsilon \big ( v_{\nu} ^{\alpha}\tilde{\nu}^{\alpha}\big)  =0
$. Here $\tilde{\nu}^\alpha(\epsilon,\chi)$  is the valley-dependent density of states, Eq.~\eqref{eq:global_DOS}, and  $
v^{\alpha}_{\nu} (\epsilon, \chi)= 2 e V (t) \mathds{V}^{\alpha}(\epsilon, \chi)
$, is the level ``velocity'' in energy space, with the sensitivity $\mathds{V}^{\alpha}(\epsilon, \chi)$ given by Eq.~\eqref{eq:level_velocity}.
The left-hand side of Eq.~\eqref{EQ:N_DOT} describes the evolution of the distribution function due to entrainment of the quasiparticles by the spectral flow, while the right-hand side describes relaxation due to inter-valley and inelastic scattering. 
Equation~\eqref{EQ:CURRENTEUL} reflects the fact that the contribution of each occupied quasiparticle state to the current is equal to $ e\mathds{V}^\alpha(\epsilon, \chi)$~\cite{liu2023current}.

The density of states in the normal region of diffusive SNS junctions differs from that of a normal metal only in the energy interval $0<\epsilon \lesssim E_T $, where $E_T\sim D/L^2$ is the Thouless energy of the junction. For our purposes we need only rough features of the energy and phase-dependence of the global density of states. For $\tau_v E_T\gg 1 $ these features can be understood by solving the independent Usadel equations for each valley, and will have the same form as that for a single valley case \cite{zhou1998density},
\begin{equation}
\label{eq:DensityOfStates}
\tilde \nu^\alpha(\epsilon, \chi) =    \nu^\alpha_N\mathcal{V} \begin{cases}
h^\alpha(\epsilon, \chi),   & \epsilon  \ \lesssim \ E_T, \\
1,  & \epsilon \gg E_T.
\end{cases}
\end{equation}
Here $\mathcal{V}$ is the volume of the normal region, and $h^\alpha(\epsilon, \chi)$ are valley-dependent functions of order unity, which are periodic in $\chi$.

\subsection{ I-V characteristics  of multivalley SNS junctions\label{sec:I-V}}

We now apply the zero-dimensional description of Sec.~\ref{sec:zero_dimension} to obtain the I-V characteristics of multivalley SNS junctions in the low bias regime. 
Generally, the time-dependence of the 
voltage $V(t)$ and the current through the junction $I(t)$ is determined by  the external circuit. We will consider two common setups; constant voltage bias, and constant current bias, and  will focus on the time-averaged voltage, $V=\overline{V(t)}$,  and current, $I=\overline{I(t)}$.

\subsubsection{Voltage bias}
\label{sec:voltage-bias}

It is shown below that the I-V characteristic of voltage-biased junctions turns out to be nonmonotonic, with two pronounced peaks of similar height $I_\text{max,1} \sim I_\text{max,2} $, as illustrated in Fig.~\ref{fig:VoltageBiasIV}.  The first peak is reached at $eV\sim \tau_{in}^{-1}$, and the second at $eV\sim \tau_{v}^{-1}$. 
We will show that the peak currents $I_\text{max,1} \sim I_\text{max,2} $ can be significantly larger than the temperature-dependent critical current $I_{c}(T)$, and in some cases may be as large as the zero temperature critical current, $I_{c}(0)$. At even larger voltages the I-V characteristic reaches a minimum, after which the device crosses over to the high bias regime where the nonlinear conductance becomes of the order of the normal state conductance $G_N$.

To obtain a quantitative description of voltage-biased junctions it is convenient to change the independent variable from the energy $\epsilon$ to the integrated density of states,
\begin{equation}\label{eq:Nintegrated}
N^\alpha(\epsilon,\chi) = \int^\epsilon_0 d \epsilon' \tilde{\nu}^\alpha(\epsilon',\chi). 
\end{equation}
The change of variables $\epsilon \to N^\alpha(\epsilon,t)$ is analogous to a transformation from Eulerian to Lagrangian representation in fluid mechanics. In this analogy, the quasiparticle energy levels correspond to the particles of a liquid, positioned at the Eulerian coordinate $\epsilon$ in energy space. The Lagrangian coordinate $N^\alpha (\epsilon)$ corresponds to the level number in valley $\alpha$ counted from zero energy. 
The density of states,  $\tilde{\nu}^\alpha(\epsilon,\chi) = \frac{\partial N^\alpha (\epsilon,\chi)}{\partial \epsilon} $,  corresponds to the density of the liquid.

In Lagrangian variables, the kinetic equation Eq.~\eqref{EQ:N_DOT} takes the simple form,
\begin{equation}
\label{eq:f_dot}
\frac{\partial}{\partial t}f_0^\alpha(N^\alpha, t) =  \text{St}^\alpha_{\mathrm{v}}  + \text{St}^\alpha_\text{in}.
\end{equation} 
The level sensitivity $\mathds{V}^\alpha(\epsilon,\chi)$ is expressed as,
\begin{equation}
 \mathds{V}^\alpha(\epsilon,\chi) = - \frac{1}{\tilde{\nu}^\alpha(\epsilon,\chi)}
 \frac{\partial N^\alpha(\epsilon, \chi) }{\partial \chi},
\end{equation}
and the current in Eq.~\eqref{EQ:CURRENTEUL} acquires the form,
\begin{equation}\label{eq:current-Lagrangian}
I = I_c(\chi,T) + \sum_\alpha e \int_0^\infty dN^\alpha  \delta f^\alpha_0 \frac{\partial \epsilon}{\partial \chi} .
\end{equation}
In this expression  the arguments of $\delta f_0^\alpha(N^\alpha,t)$ and $\epsilon(N^\alpha,\chi)$ are suppressed. 

In the voltage interval  $e V \ll 1/\sqrt{\tau_{in}\tau_v}$ we may neglect the inter-valley relaxation in Eq.~\eqref{eq:f_dot}, and obtain its solution
 in the general form,  
\begin{equation}\label{eq:nGenE}f_0^\alpha(N^\alpha,t) = \int_0^\infty\frac{ d\chi' e^{\frac{-\chi'}{2eV\tau_{in}}}}{2eV\tau_{in}} f_F[\epsilon(N^\alpha, \chi(t)-\chi')].
\end{equation}
Using Eqs.~\eqref{eq:DensityOfStates} and \eqref{eq:nGenE} (Appendix~\ref{nonlinearConductance} for details)
the nonlinear conductance $G(V)=I(V)/V$ may be estimated from these expressions for $T \gg E_T$ as, 
\begin{equation}
\label{eq:GaverageD}
\frac{G(V)}{G_N}  \sim \frac{E_T^2}{T}   \begin{cases}
\tau_{in} &    eV\ll \frac{1}{\tau_{in}}, \\
 \frac{\tau_{in}}{(2eV\tau_{in})^{2}} &  \frac{1}{\tau_{in}} \ll eV \ll \frac{1}{\sqrt{\tau_{in}\tau_v}}.
 \end{cases}
\end{equation}
Here $G_N \sim e^2 \sum_\alpha \nu_N^{\alpha} \mathcal{V} E_T$ is the conductance of the junction in the normal state. It follows from  Eq.~\eqref{eq:GaverageD} that at  $eV_1 \sim 1/\tau_{in}$ the current has a maximum of magnitude 
\begin{equation}\label{eq:JmaxAv}
    I_\text{max,1} \sim I_c(0)\frac{E_T}{T},
\end{equation}
where $I_c(0) \sim E_TG_N/e$ is the critical current of a diffusive SNS junction at $T=0$. We note that $I_\text{max,1}$ may be significantly larger than the finite temperature critical current $I_c(T)$.

For $e V \gg 1/\sqrt{\tau_{in}\tau_v}$ we may neglect the inelastic relaxation in Eq.~\eqref{eq:f_dot}. The nonlinear conductance in this regime is described by Eq.~\eqref{eq:GaverageD} with $\tau_{in}$ replaced by $\tau_v$  and the voltage ranges replaced by,  $1/\sqrt{\tau_{in}\tau_v}\ll eV\ll 1/\tau_v$ and $1/\tau_v\ll eV\ll E_T/\sqrt{\tau_v T}$, \footnote{The upper limit in voltage is determined by the condition $G(V)/G_N \sim 1$.}. Thus, a second current peak of magnitude $I_{\mathrm{max},2} \sim I_{\mathrm{max},1} $ appears at a bias voltage $e V_2 \sim 1/\tau_v$.

At even larger voltages the dominant contribution to the current is due to the normal state contribution,  which is a monotonically increasing function of voltage, $I \sim G_NV$. The resulting I-V characteristic at voltage bias is summarized in Fig.~\ref{fig:VoltageBiasIV}. 

\subsubsection{Current bias}
\label{sec:current-bias}

\begin{figure*}
    \centering
\includegraphics[width=.495\linewidth]{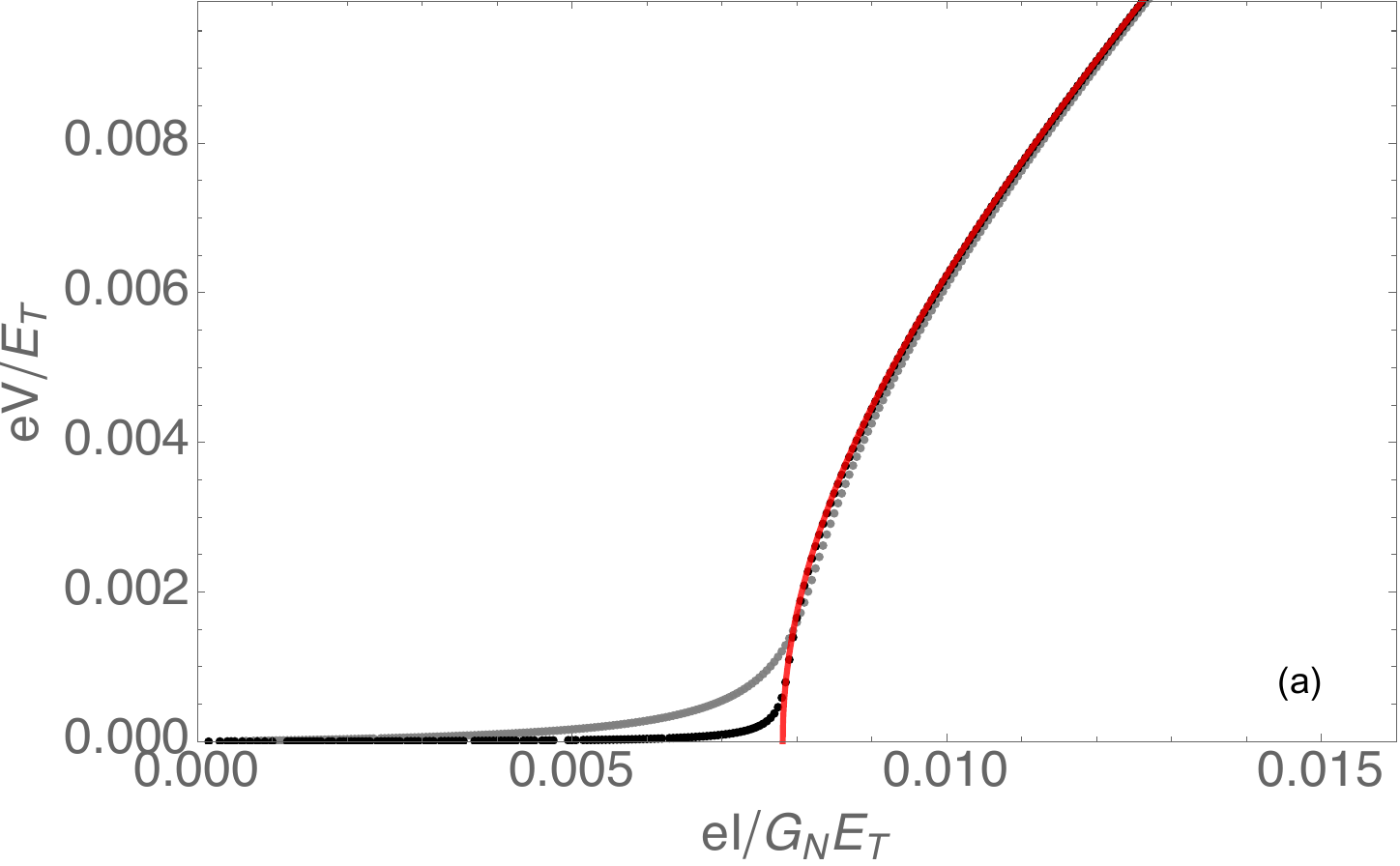} \includegraphics[width=.495\linewidth]{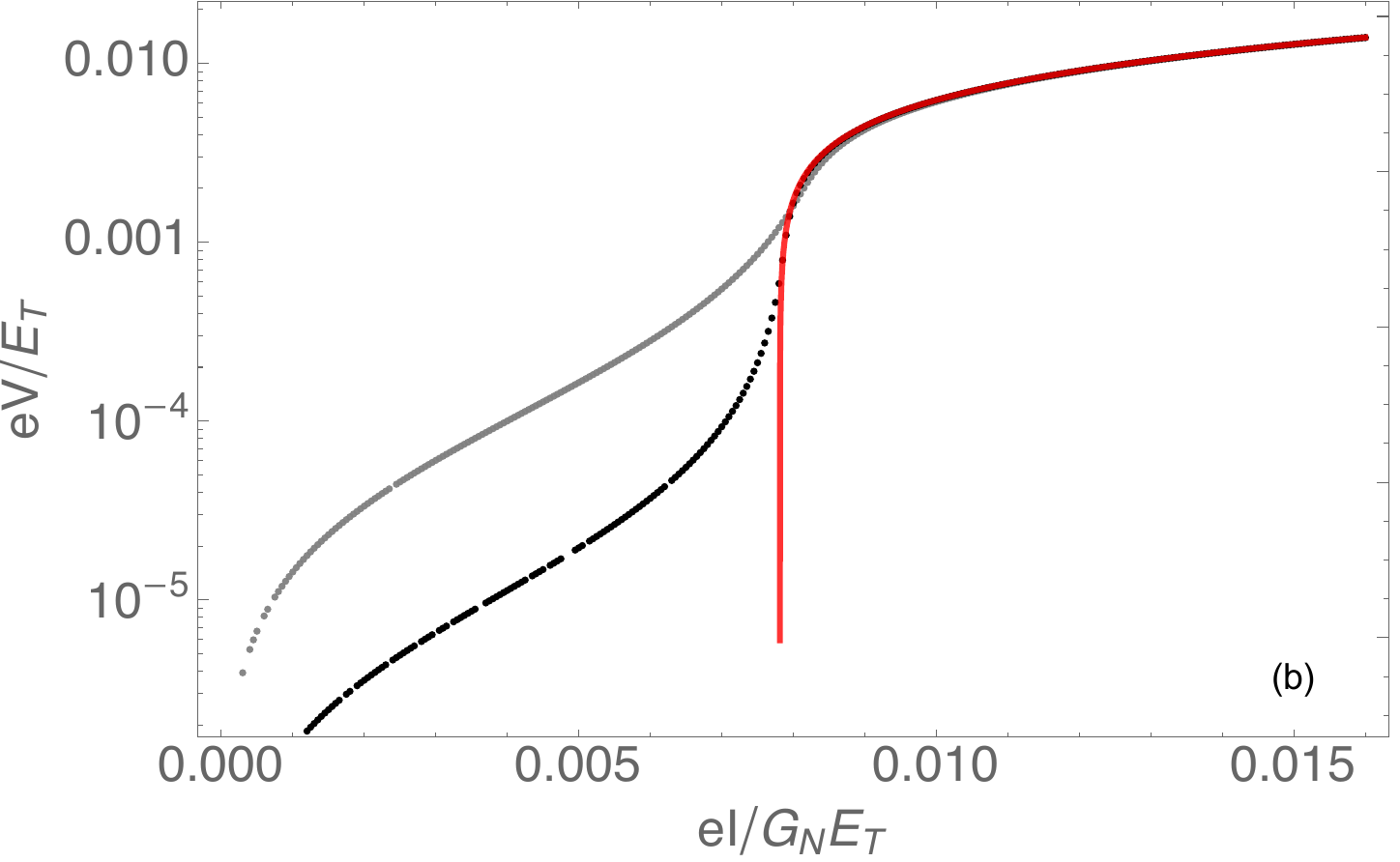}
     \caption{I-V characteristics of current-biased SNS junctions obtained by solving the model in Appendix~\ref{sec:model}. (a) Plot of dimensionless voltage $eV/E_T$ versus dimensionless current $eI/G_NE_T$ for $T/E_T=4$. Black dots are the solutions to Eqs.~\eqref{eq:delta_n_ODE} and~\eqref{eq:avgvoltage} with $1/(\tau_{in} E_T) =10^{-5} $ and the gray dots are the solutions with $1/(\tau_{in} E_T) =10^{-4}$. The high current asymptotic, Eq.~\eqref{eq:highcurrent} is plotted in red. The jump current is $I_{\text{jump}} = G_NE^2_T/32eT$, which for the chosen parameters occurs at $eI_{\text{jump}}/G_NE_T\approx 0.0079$. (b) The same plot on a log-linear scale. At bias currents below $I_\text{jump}$ the nonlinear conductance is proportional to $\tau_{in}$. This is illustrated in plot (b) where black and gray traces are separated by a factor of ten on the log-scale.} \label{fig:CurrentBiasIV}
\end{figure*}

It follows from  Eq.~\eqref{eq:DensityOfStates} that only the quasiparticles in energy levels within a narrow window $\epsilon \sim E_T \ll T$  contribute to the current (below, we refer to them as ``active"  levels). This observation enables us to introduce a simple model of the junction, which describes the I-V characteristics of the junction  with accuracy of order unity. 

In this model, the sensitivities of the quasiparticle levels in the active group (which are of the same order of magnitude) are replaced by the typical sensitivity. 
As a result, the nonequilibrium occupancy of all quasiparticle levels within the active group is replaced by the typical occupancy, $\delta n$.
Therefore, the kinetic equation  for the distribution function reduces to a differential equation for the occupancy of the single active level. The current across the junction is expressed within this model in terms of the sensitivity of the active level and its occupancy. This model can be solved numerically under constant current bias.

We find that for generic model parameters, the hierarchy of relaxation times does not produce additional features in the  I-V characteristic. We therefore relegate the details of the model and its solution to Appendix~\ref{sec:model}, and plot the results of the numerical solution in Fig.~\ref{fig:CurrentBiasIV}.  
At low current bias, the I-V characteristic exhibits a distinct low voltage regime, where the nonlinear conductance  is controlled by the long inelastic relaxation time $\tau_{in}$. At high current bias the nonlinear conductance approaches the conductance of the junction in its normal state, $G_N$. The transition between these two regimes is sharp and occurs at a bias current, $I_\text{jump}\sim I_\text{max,1}$~\cite{liu2023current}. 
The model in Appendix~\ref{sec:model} enables us to describe the shape of the transition from the low-bias to high-bias regime.

It is shown in Sec.~\ref{sec:high_bias} that even in the regime of relatively high voltage, $eV \gg 1/\tau_{in}$, the nonequilibrium quasiparticle population in the active levels provides a significant contribution to the current, as illustrated by the red curves in Fig.~\ref{fig:CurrentBiasIV}.

\section{Summary}

We generalized the Larkin-Ovchinnikov equations for disordered nonequilibrium superconductors to multivalley materials. The system of equations consists of the generalized Usadel equation Eq.~\eqref{eq:multiUsadel} for the retarded/advanced Green's functions, the kinetic equations Eqs.~(\ref{eq:kineq0}-\ref{eq:CollisionIntegrals}) for the quasiparticle distribution functions $f^\alpha_0(\epsilon)$ and $f^\alpha_1(\epsilon)$, and Eqs.~(\ref{eq:currentdensity}-\ref{eq:self-consistent}) which express the electron density, 
current density, and the order parameter in terms of the quasiparticle distribution functions. 

In the presence of the electric field, the density of quasiparticle states in superconductors becomes time-dependent. We use the identity Eq.~\eqref{eq:conservation dos} to write the generalized LO equations \eqref{eq:kineq0} and \eqref{eq:kineq1} in a form that explicitly contains the time-dependent density of states.

Using this generalization, in Sec.~\ref{sec:SNS} we studied the I-V characteristics of SNS junctions with a multivalley normal region. For voltage-biased junctions considered in Sec.~\ref{sec:voltage-bias}, the I-V characteristic is nonmonotonic. The  current across the junction exhibits two sharp peaks of similar height $I_\text{max,1}, \ I_\text{max,2}$, which are parametrically larger than the critical current $I_c$. The peaks are located at voltage biases that correspond to the relaxation rates of the inter-valley and inelastic relaxation. 
For current-biased junctions considered in Sec.~\ref{sec:current-bias}, the I-V characteristic exhibits a step-like increase in the voltage as the bias current becomes of order $I_\text{max,1}$. Above $I_\text{max,1}$ the nonlinear resistance quickly approaches that of the normal state of the junction, $G_N^{-1}$. In a wide interval of bias currents $I_c< I <I_\text{max,1}$, the nonlinear conductance of the junction parametrically exceeds $G_N$ and is controlled by the inelastic and inter-valley relaxation times. 

\section*{Acknowledgments}
The authors are grateful to C. Marcus for useful discussions. The work of B.S., S-B.C., and L.B. was supported by the
DARPA grant GR049687, A.A. was supported by the
NSF grant DMR-2424364.

\appendix
\begin{widetext}
\section{\label{Appendix}Derivation of the kinetic equations for diffusive multivalley superconductors}

The disorder-averaged matrix Green's function $\left<\check{{G}}^\alpha\right>$ obeys the Dyson equation,
\begin{equation}
    \left[i\check \tau_3 \vv{\partial}_t - \check H^\alpha_\text{BCS}- \check{\Sigma}^\alpha\right]\circ\left<\check{{G}}^\alpha\right> - \left<\check{{G}}^\alpha\right>\circ\left[{i\check \tau_3 \backvv{\partial}_{t'} }-\check H^\alpha_\text{BCS} - \check{\Sigma}^\alpha\right] =0,
    \label{eq:dyson}
\end{equation}
where the BCS Hamiltonian $\check{H}_{BCS}^\alpha$ was defined in Eq.~\eqref{eq:H_BdG}, and $\circ$ denotes the product of operators. The self-energy, $\check{\Sigma}^\alpha$, describes scattering by the  disorder potentials $U^\alpha$ and $v^{\alpha\beta}$, as well as inelastic scattering, 
\begin{eqnarray}
    \check{\Sigma}^\alpha = \check{\Sigma}_{\text{intra}}^\alpha + \check{\Sigma}_{\text{inter}}^\alpha +\check{\Sigma}_{\text{in}}^\alpha.
\end{eqnarray}For white noise disorder, the elastic contributions to the self-energy may be expressed in terms of the Green's function at coinciding points, 
\begin{equation}
    \label{eq:self-energy_intra}
\check{\Sigma}_\text{intra}^\alpha(\mathbf{x},t_1,t_2) = \frac{1}{2\pi\tau \nu_N^\alpha} \left<\check{{G}}^\alpha(\mathbf{x},t_1,\mathbf{x},t_2)\right>, \quad \check{\Sigma}_\text{inter}^\alpha(\mathbf{x},t_1,t_2) = \sum_{\beta \neq \alpha} \frac{1}{2\pi  \tau_{\alpha\beta}  \nu_N^\beta} \left<\check{G}^\beta(\mathbf{x},t_1,\mathbf{x},t_2)\right>,
\end{equation} 
where the rates of intra- and inter-valley relaxation are expressed in terms of the variance of the disorder potentials $U^\alpha$ and $v^{\alpha\beta}$ as, 
 \begin{equation}\label{eq:variances}
     \nu_N^\alpha \left<U^\alpha (\mathbf{x}_1)U^\alpha (\mathbf{x}_2)\right> = \frac{1}{2\pi\tau } \, \delta(\mathbf{x}_1-\mathbf{x}_2), \quad  \nu_N^\gamma \left<v^{\alpha\gamma} (\mathbf{x}_1)v^{ \gamma \alpha} (\mathbf{x}_2)\right> = \frac{1}{2\pi  \tau_{\alpha\gamma}  }\, \delta(\mathbf{x}_1-\mathbf{x}_2). 
\end{equation} 
No summation over the repeated indices should be performed in these expressions.

The quasiclassical Green's functions in the Larkin-Ovchinnikov approach are obtained as follows. One introduces the center-of-mass coordinates $(t,\mathbf{r})\equiv \left(\mathbf{X}_1+\mathbf{X}_2\right)/2$ and relative coordinates $(t',\mathbf{r}')\equiv\left(\mathbf{X}_1-\mathbf{X}_2\right)$. Next, one performs the Fourier transform over the relative coordinate $(t',\mathbf{r}')\rightarrow(\epsilon,\mathbf{p})$ to obtain Green functions in the Wigner representation. Then, integrating over $\xi^\alpha_p = \frac{p^2}{2m}-\mu^\alpha$ defines the quasiclassical Green's function, 
\begin{equation}
\check{g}^\alpha\left(\epsilon,t,\mathbf{r},\mathbf{n}\right) = \frac{i}{\pi}\int d\xi^\alpha_p \int dt'd^3r'\left(e^{i\epsilon t' - i\mathbf{p}\cdot\mathbf{r}'}\left<\check{{G}}^\alpha(\mathbf{X}_1,\mathbf{X}_2)\right>\right).\label{eq:quasiclassicalgf}
\end{equation}
Here, $\mathbf{n}$ denotes the direction of the momentum $\mathbf{p}$ conjugate to the relative coordinate $\mathbf{r}'$. The quasiclassical Green's function obeys the nonlinear constraint, $(\check{g}^\alpha)^2=1$~\cite{larkin1977non}. Linearizing $\check H_\text{BCS}^\alpha$ around the Fermi surface and integrating Eq.~\eqref{eq:dyson} over $\xi^\alpha_p$, one obtains the Eilenberger equation, 
\begin{equation}
\frac{1}{2}\partial_t\left\{\check{\tau}_3,\check{g}^\alpha\right\}+\left (v_F^\alpha\right)_i\left(\partial_i\check{g}^\alpha -\frac{e}{2}\partial_t A_i \left\{\check{\tau}_3,\partial_\epsilon\check{g}^\alpha\right\}\right) -\frac{1}{2}\left\{\partial_t\check{\Delta}^\alpha,\partial_\epsilon\check{g}^\alpha\right\}+\partial_t(e\phi - \mu^\alpha)\partial_\epsilon\check{g}^\alpha-i\left[\epsilon\check{\tau}_3 +\check{\Delta}^\alpha +\check{\Sigma}^\alpha,\check{g}^\alpha\right] =0. \label{eq:eilenberger}
\end{equation}
The covariant derivative here is defined as $\partial_j = \nabla_j - ie A_j\left[\check{\tau}_3,\cdot\right]$,  and $(v^\alpha_F)_i$ is the Fermi velocity along the $i$-axis of valley $\alpha$. To obtain Eq.~\eqref{eq:eilenberger} the products of operators in the Wigner representations were evaluated to first order accuracy in the Moyal expansion \cite{moyal1949quantum,groenewold1946principles,baker1958formulation},
\begin{equation}
    \check{A}\circ \check{B} \approx  \check{A} \left[1+\frac{i}{2}\left(\backvv{\partial}_x \vv{\partial}_p - \backvv{\partial}_p \vv{\partial}_x -\backvv{\partial}_t \vv{\partial}_\epsilon + \backvv{\partial}_\epsilon \vv{\partial}_t\right)\right] \check{B}.
\end{equation}
The self-energies due to intra- and inter-valley scattering, defined in Eq.~\eqref{eq:self-energy_intra}, are given by 
\begin{equation}\label{eq:selfenergyQC}
    \check\Sigma^\alpha_\text{intra} = \frac{-i}{2\tau}\left<\check{g}^\alpha(\epsilon,t,\mathbf{r},\mathbf{n})\right>_\mathbf{n}, \quad  \check\Sigma^\alpha_\text{inter} = \sum_{\beta \neq \alpha}\frac{-i}{2\tau_{\alpha\beta}}\left<\check{g}^\beta(\epsilon,t,\mathbf{r},\mathbf{n})\right>_\mathbf{n},
\end{equation}
in the quasiclassical regime, where averaging over the direction $\mathbf{n}$ is denoted by $\left<\cdot\right>_\mathbf{n}$.

In the diffusive regime,  the $\mathbf{n}$-dependence of  Green's functions is expanded to first order in the angular harmonics,
$\check{g}^\alpha\left(\epsilon,t,\mathbf{r},\mathbf{n}\right) = \check{g}_0^\alpha\left(\epsilon,t,\mathbf{r}\right)+\mathbf{\check{g}}_1^\alpha\left(\epsilon,t,\mathbf{r}\right) \cdot \mathbf{n}$ with $\mathbf{\check{g}}_1^\alpha\cdot \mathbf{n} \ll \check{g}^\alpha$. By employing the normalization condition for the quasiclassical Green's function, the amplitude of the first angular harmonic, $\mathbf{\check{g}}_1^\alpha$, can be expressed in terms of the gradient of the zeroth angular harmonic as,
 \begin{equation}
   \left(\check{g}_1^\alpha\right)_i = -\tau\left(v_F^\alpha\right)_i \check{g}_0^\alpha\partial_i\check{g}_0^\alpha.
 \end{equation}
Substituting $\check{g}^\alpha$ back into the Eilenberger equation \eqref{eq:eilenberger}  and averaging over directions $\mathbf{n}$, one obtains the multivalley generalization of the Larkin-Ovchinnikov equations for $\check{g}^\alpha$,  
\begin{align}
    \frac{1}{2}\partial_t\left\{\check{\tau}_3,\check{g}^\alpha\right\}-D^\alpha_{ij}\left[\partial_i\left(\check{g}^\alpha\partial_j\check{g}^\alpha \right) - \frac{e}{2}\partial_tA_i\left\{\check{\tau}_3,\partial_\epsilon\left(\check{g}^\alpha\partial_j\check{g}^\alpha\right)\right\}\right] - i\left[\epsilon\check{\tau}_3+\check{\Delta}^\alpha+\check{\Sigma}_\text{in}+\sum_\beta\frac{i}{2\tau_{\alpha\beta}}\check{g}^\beta,\check{g}^\alpha\right]&\nonumber\\-\frac{1}{2}\left\{\partial_t\check{\Delta}^\alpha,\partial_\epsilon\check{g}^\alpha\right\}+\partial_t(e\phi-\mu^\alpha)\partial_\epsilon\check{g}^\alpha=&0. \label{LarkinOvchinnikov}
\end{align}
Here we defined the valley-dependent diffusion tensor   $D^\alpha_{ij}=\tau\left<(v^\alpha_F)_i(v^\alpha_F)_j\right>_\mathbf{n}$, and for the sake of brevity 
suppressed the obvious arguments in the Green's functions, $\check{g}^\alpha_0(\epsilon,t,\mathbf{r})\rightarrow \check{g}^\alpha$.

In the adiabatic regime, where the parameters $\Delta(t)$, and $\mathbf{A}(t)$ of the BCS Hamiltonian Eq.~\eqref{eq:H_BdG} vary slowly with time, the retarded/advanced components of Eq.~\eqref{LarkinOvchinnikov} are automatically satisfied by 
the solution of the multivalley Usadel equation Eq.~\eqref{eq:multiUsadel}, 
$\hat{g}_R^\alpha\left(\epsilon,\mathbf{r},\Delta(t),\mathbf{A}(t)\right)$. Taking the difference of the retarded and advanced components of Eq.~\eqref{LarkinOvchinnikov} and evaluating the trace over the Gor'kov-Nambu space, one obtains Eq.~\eqref{eq:conservation dos}, which expresses the conservation law for the local density of states in valley $\alpha$. Similarly,  multiplying the difference of the retarded and advanced components of Eq.~\eqref{LarkinOvchinnikov} by $\hat\tau_3$ and taking the trace over the Gor'kov-Nambu space, one obtains the following relation for the spectral current density $j_j^\alpha$, defined in Eq.~\eqref{eq:dosDefinition}, 
\begin{equation}
     D^\alpha_{ij}\partial_i j^\alpha_j = \sum_{\beta \neq \alpha}\frac{1}{2\tau_{\alpha\beta}}\Tr[(\hat g_R^\beta \hat g_R^\alpha - \hat g_A^\beta \hat g_A^\alpha) \hat\tau_3 - (\alpha \leftrightarrow \beta)\hat \tau_3]\label{conservation2}.
\end{equation}
Summing this expression over all valleys one obtains the relation
$\sum_{\alpha} D_{ij}^\alpha \partial_ij_j^\alpha = 0$, which reflects the fact that inter-valley scattering does not change the  total spectral current.

The time-evolution of quasiparticle distribution functions is described by the Keldysh component of Eq.~\eqref{LarkinOvchinnikov}. The normalization Eq.~\eqref{eq:nonlinearconstraint} of the Green's function $\check{g}^\alpha$ allows the Keldysh component to be parameterized as,
$\hat{g}^\alpha_K = \hat{g}^\alpha_R \hat{h}^\alpha -\hat{h}^\alpha\hat{g}^\alpha_A$, for any matrix $\hat{h}^\alpha$. It has been shown that there are only two independent components of $\hat h^\alpha$ in the semiclassical approximation \cite{kopnin2001theory}. Following the Larkin-Ovchinnikov parameterization \cite{larkin1977non}, we take $\hat{h}^\alpha = f^\alpha_0\hat{\tau}_0 + f^\alpha_1\hat{\tau}_3$ where $f^\alpha_{0},f_1^\alpha$ are odd and even functions of $\epsilon$ respectively. In equilibrium $f^\alpha_0 = \tanh(\epsilon/2T)$ and $f^\alpha_1 = 0$. 
Taking the trace of the Keldysh component of Eq.~\eqref{LarkinOvchinnikov} multiplied by $\hat\tau_0$ and $\hat\tau_3$, and making use of Eqs.~\eqref{eq:conservation dos} and \eqref{conservation2},
we  obtain the kinetic equations for $f^\alpha_0$ and $f^\alpha_1$, Eqs.~(\ref{eq:kineq0}-\ref{eq:CollisionIntegrals}).

The time evolution equations for the Green's functions contain the self-consistent fields, $\Delta,\phi,A_i$. These fields are related to the Green's functions by the BCS self-consistency equation, the continuity equation for the electron density, and the Maxwell equations. In metals, the Poisson equation yields the charge neutrality condition, Eq.~\eqref{eq:charge_neutrality}. In multivalley superconductors this condition may be satisfied for nonvanishing deviations of the electron density in a given valley from equilibrium, $\delta n^\alpha\neq 0$. The continuity equation for the electron density,
\begin{equation}\label{eq:chargedensity}
    \delta \dot{ n}^\alpha + \frac{1}{e}\partial_i J^\alpha_i + \mathcal{I}^\alpha_\Delta + \mathcal{I}^\alpha_\text{impurity} = 0,
\end{equation}
is obtained by multiplying the Keldysh component of Eq.~\eqref{LarkinOvchinnikov} with $\hat \tau_3 \nu^\alpha_N/4$, taking the trace, and integrating over the energy. Here, the electron density $n^\alpha$ and current density $J_i^\alpha$ in valley $\alpha$ are defined, respectively, by Eqs.~\eqref{eq:electron-density} and \eqref{eq:currentdensity}. Finally, 
\begin{equation}
\begin{split}
\mathcal{I}^\alpha_\Delta = -i\frac{\nu_N^\alpha}{2}\int^\infty_0 d\epsilon\left(\Tr \left[\hat \tau_3\hat \Delta^\alpha (\hat{g}^\alpha_R -\hat{g}^\alpha_A)\right]f_0^\alpha + \Tr \left[\hat\Delta^\alpha(\hat{g}^\alpha_R  +\hat{g}^\alpha_A)\right]f_1^\alpha\right) 
\end{split}
\end{equation} 
and
\begin{equation}
    \begin{split}
     \mathcal{I}^\alpha_\text{impurity} 
       = \frac{\nu^\alpha_N}{4} \int^\infty_0 d\epsilon\sum_{\beta\neq \alpha} \frac{1}{\tau_{\alpha \beta}}\Re\bigg[\Tr\left[(\hat{g}^\alpha_A \hat{\tau}_3\hat{g}_R^\beta)[(f^\beta_0 -f^\alpha_0) + (f^\beta_1-f^\alpha_1)\hat{\tau}_3]\right] \\+ \Tr\left[ (\hat{g}_R^\beta\hat{g}_R^\alpha)(f^\alpha_1- f^\beta_1)\right] + \Tr\left[ (\hat{g}_R^\beta\hat{g}_R^\alpha \hat \tau_3)f^\alpha_0 - (\hat{g}_R^\alpha\hat{g}_R^\beta \hat \tau_3)f^\beta_0\right]\bigg].
    \end{split}\label{influx2}
\end{equation}
describe the outflow of electrons from valley $\alpha$
due to the inter-valley BCS coupling and the inter-valley impurity scattering.

\section{Estimate of the nonlinear conductance of voltage-biased junctions}\label{nonlinearConductance}

The nonlinear conductance, Eq.~\eqref{eq:GaverageD}, may be obtained using the Lagrangian description of the current,  Eq.~\eqref{eq:current-Lagrangian} following Ref.~\onlinecite{liu2023current}. 
The deviations of the quasiparticle energies from their averages may be written in terms of the Lagrangian coordinates, 
\begin{equation}
    \delta \epsilon(N^\alpha,\chi) = \epsilon(N^\alpha,\chi) - \overline{\epsilon(N^\alpha)},
\end{equation} 
where $\overline{\epsilon(N^\alpha)}  =\int_0^{2\pi} \epsilon(N^\alpha,\chi) d \chi/2\pi $. These periodic functions may be expanded in a Fourier series in $\chi$, 
\begin{equation}
    \delta \epsilon(N^\alpha,\chi) = \sum_{k\neq 0} C_k(N^\alpha) e^{ik\chi}.
\end{equation}
For temperatures $T\gg E_T$ the equilibrium level occupancy may be expanded to linear order in $\delta \epsilon$,
\begin{equation}\label{eq:fdeltaE}
f_F(\epsilon(N^\alpha,\chi(t)))  \approx f_F[\overline{\epsilon(N^\alpha)}] + \partial_\epsilon f_F[\overline{\epsilon(N^\alpha)}] \delta \epsilon(N^\alpha,\chi(t)).
\end{equation}
Substituting Eq.~\eqref{eq:fdeltaE} into  Eq.~\eqref{eq:nGenE} we obtain for $eV \ll 1/\sqrt{\tau_v\tau_{in}}$,
\begin{equation}\label{eq:f(t)deltaE}
    f_0^\alpha(N^\alpha,t) = f_F[\overline{\epsilon(N^\alpha)}] + \frac{\partial_\epsilon f_F[\overline{\epsilon(N^\alpha)}]}{2e\tau_{in}V}\sum_{k\neq 0} \frac{C_k(N^\alpha) e^{ik\chi}}{ik+(2eV\tau_{in})^{-1}}.
\end{equation}
The current Eq.~\eqref{eq:current-Lagrangian} averaged over the period of phase winding may then be expressed as,
\begin{equation}
    I(V) = e\sum_\alpha \int_0^\infty d N^\alpha \partial_\epsilon f_F[\overline{\epsilon(N^\alpha)}] \sum_{q,k\neq 0 } \left<\frac{iqe^{i(k+q)\chi}C_k(N^\alpha)C_q(N^\alpha)}{2e\tau_{in}V(ik)+1}\right>_\chi.
\end{equation}
 Noting that the Fourier components of a real function satisfy $C_k=C_{-k}^*$ we obtain, 
\begin{equation}
\label{eq:CurrentCk}
I(V) = e\sum_\alpha \int_0^\infty d N^\alpha \partial_\epsilon f_F[\overline{\epsilon(N^\alpha)}] \sum_{k\neq 0 } \frac{2eV\tau_{in}k^2\abs{C_k(N^\alpha)}^2}{1+(2e\tau_{in}Vk)^2}.
\end{equation}
The estimate of the nonlinear conductance  may be obtained from this expression using Eq.~\eqref{eq:DensityOfStates}. The quasiparticle levels, which have an appreciable  sensitivity to the phase difference $\chi$, lie within the energy window $0< \epsilon \lesssim E_T$. The number of such levels may be estimated as $\sum_\alpha E_T (\mathcal{V} \nu_N^\alpha) \sim G_N/e^2$, where $G_N$ is the  conductance of the junction in the normal state. 
For such levels, the leading Fourier amplitudes have $k\sim 1$, and may be estimated as $C_k (N^\alpha) \sim E_T$. Substituting these estimates into Eq.~\eqref{eq:CurrentCk} we obtain Eq.~\eqref{eq:GaverageD}. 

\section{Solvable model of current-biased junction \label{sec:model}}

The quasiparticle levels in the ``active" group ($\epsilon \lesssim E_T$) have similar sensitivity to phase changes.
At  $T\gg E_T$, these levels have similar relaxation rates and thus similar occupancy. 
The level occupancy $n$ satisfies the kinetic equation,
\begin{equation}\label{Eq:dotnlevel}
   \dot{n} = -\frac{n-n_F(\epsilon(\chi))}{\tau_{in}} .
\end{equation}
We denote the deviation of the typical occupancy  from the equilibrium 
distribution by $\delta n$, and the typical sensitivity by $\epsilon'(\chi)=\partial_\chi \epsilon(\chi)$. Expanding $n_F(\epsilon(\chi))$ to first order in $\epsilon(\chi)/T$ we get,
\begin{equation}\label{eq:n_dot}
  \delta \dot{ n} - \frac{\epsilon' (\chi)}{4 T} \dot{\chi} = -\frac{\delta n}{\tau_{in}}.
\end{equation}

The current across the junction 
may be expressed in terms of the number of quasiparticle levels in the active group, $N_a$, and the nonequilibrium occupancy $\delta n$ in the form, 
\begin{equation}\label{eq:current}
     I= eN_a \epsilon'(\chi) \delta n(\chi) + \frac{G_N\dot{\chi}}{2e}.
\end{equation}
Here we neglected the equilibrium current as the contribution of nonequilibrium quasiparticles to the current can greatly exceed it. This assumption restricts the applicability of the model to bias currents much greater than the temperature-dependent critical current, $I\gg I_c(T)$. The last term on the right-hand side of Eq.~\eqref{eq:current} represents the contribution to the current arising from the distribution function $f^\alpha_1$. At small voltages this contribution may be neglected, while at large voltages it corresponds to the normal state current $G_N \dot{\chi}/2e$. Thus, the expression Eq.~\eqref{eq:current} should be understood as an interpolation formula between the low bias and the high bias regimes.

The number of quasiparticle levels in the active group  may be estimated as $N_a \sim \mathcal{V}E_T\sum_\alpha \nu^\alpha_N \sim G_N/e^2$, where $(\nu^\alpha_N\mathcal{V})^{-1}$ is the mean level spacing. To reduce the number of parameters in the model, in the expressions below we set $N_a = G_N/e^2$. The system of equations  \eqref{eq:n_dot} and \eqref{eq:current} is solved as follows. Since the phase $\chi$ increases monotonically with time, the level occupancy may be viewed as a periodic function $\delta n(\chi)$. The differential equation for $\delta n(\chi)$ is obtained by excluding the time differential from Eqs.~\eqref{eq:n_dot} and \eqref{eq:current}.  From Eq.~\eqref{eq:n_dot} one gets, 
\begin{equation}\label{eq:chi_dot}
    d\ln(\delta n) -\frac{1}{4T}\frac{d\epsilon}{\delta n} = -\frac{dt}{\tau_{in}}.
\end{equation}
On the other hand,   Eq.~\eqref{eq:current} yields
\begin{align}\label{eq:dt_current}
  d t = & \, \frac{(G_N/2e) d \chi}{I - (G_N/e)\epsilon'(\chi) \delta n(\chi)  }.
\end{align}
Substituting Eq.~\eqref{eq:dt_current} into Eq.~\eqref{eq:chi_dot} one obtains a nonlinear first-order ODE  for $\delta n(\chi)$,
\begin{equation}\label{eq:delta_n_ODE}
   \left[1 - \frac{G_N\epsilon' \delta n}{eI} \right]
   \left[   
  \left( \ln \delta n \right)'-  \frac{ \epsilon' }{4T\delta n} \right]= - \frac{G_N}{ 2e  \tau_{in} I }.
\end{equation}
The expression for the average voltage $V(I)$ may be obtained by noting that the period of phase winding by $2\pi$ is related to the voltage by  $\mathcal{T} = \pi/eV$. Integrating  Eq.~\eqref{eq:dt_current} over this period we get
\begin{equation}\label{eq:avgvoltage}
    V = \frac{I}{G_N}\left(\int_0^{2\pi} \frac{d\chi}{2\pi}\frac{1}{1 - G_N (\epsilon'\delta n )/eI}\right)^{-1}.
\end{equation}
For a given phase-dependence of the active level energy $\epsilon(\chi)$, the system of equations \eqref{eq:delta_n_ODE} and \eqref{eq:avgvoltage} determines the I-V characteristic of the current-biased junction. This system can be solved numerically. 

The resulting I-V characteristic shows a clear separation between the low bias regime dominated by long-time relaxation, and the high bias regime where the inelastic and inter-valley relaxation become ineffective. This can be clearly seen in Fig.~\ref{fig:CurrentBiasIV}, which shows the plot of the current-voltage dependence using the simplest form of the phase-dependence of the active level energy, $\epsilon(\chi) = E_T\cos^2(\chi/2)$.

\subsection{High bias regime\label{sec:high_bias}}

It is clear from Fig.~\ref{fig:CurrentBiasIV} that  
even in the high bias regime, where the inelastic and inter-valley relaxation become ineffective, the nonequilibrium population of active quasiparticle levels significantly affects the average voltage across the current-biased junction.

In the high bias regime, $I \gg G_N/e\tau_{in}$, the solution of the model simplifies dramatically because the active level occupancy becomes time-independent. In this case the solution of Eq.~\eqref{eq:n_dot} is given by 
\begin{equation}\label{eq:dissipationless}
    \delta n(\chi) = \frac{\epsilon(\chi)}{4T} + \overline{\delta n},
\end{equation}
where $\overline{\delta n}$ is a constant. Substituting Eq.~\eqref{eq:dissipationless} into Eq.~\eqref{eq:avgvoltage} we can express the voltage at high current bias in the form,
\begin{equation}\label{eq:highcurrent}
    V = \frac{I}{G_N}\left(\int_0^{2\pi} \frac{d\chi}{2\pi}\frac{1}{1 - G_N (\overline{\delta n} + \epsilon/4T)\epsilon'/eI}\right)^{-1}.
\end{equation}
The value of $\overline{\delta n}$ may be determined by substituting Eq.~\eqref{eq:dissipationless} into Eq.~\eqref{eq:n_dot}, and neglecting the relaxation,  $\tau_{in} \to \infty$. Integrating the result over the period, we get
the condition
\begin{equation}
    \overline{\delta n} =-\int_0^\mathcal{T} \frac{dt}{\mathcal{T}} \frac{\epsilon(\chi(t))}{4T}.
\end{equation}
Replacing the averaging over time by averaging over the phase with the aid of Eq.~\eqref{eq:dt_current} we can express  $\overline{\delta n}$
in terms of the bias current using definite integrals, 
\begin{equation}\label{eq:deltanbar}
    \overline{\delta n} = -\frac{\int_0^{2\pi} \frac{d\chi}{2\pi} \frac{\epsilon(\chi)}{4T}\left[eI-G_N\epsilon'(\chi)(\frac{\epsilon(\chi)}{4T}+\overline{\delta n})\right]^{-1}}{\int_0^{2\pi} \frac{d\varphi}{2\pi} \left[eI-G_N\epsilon'(\varphi)(\frac{\epsilon(\varphi)}{4T} +\overline{\delta n})\right]^{-1}}.
\end{equation}

For the model $\epsilon(\chi)=E_T\cos^2(\chi/2)$, the I-V characteristic obtained by this procedure  is plotted in red in Fig.~\ref{fig:CurrentBiasIV}. In the high bias regime it reproduces the result of the numerical solution of Eq.~\eqref{eq:delta_n_ODE}. The intercept of the high-bias solution with the horizontal axis occurs at the bias current $I_\text{jump}= G_NE^2_T/32eT$, which separates the high bias and the low bias regime. As expected, $I_\text{jump}$ is of the same order of magnitude as the height of the current peak $I_\text{max,1}$ at constant voltage bias. 

\end{widetext}
\bibliographystyle{apsrev4-2}
\bibliography{ref} 
\end{document}